Research Article:

# Comparative Analysis of Deep Learning Architectures for Multi-Disease Classification of Single-Label Chest X-rays

Ali M. Bahram[1], Saman Muhammad Omer[2], Hardi M. Mohammed[1]

[1]Department of Computer Science, College of Science, Charmo University, 46023 Chamchamal, Kurdistan Region, Iraq
[2]Department of Software Engineering, University of Raparin, Ranya, Kurdistan Region, Iraq
https://doi.org/10.31530/cjnst.2026.2.1.2

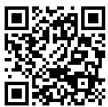

**Abstract**

**Background**: Chest X-ray imaging has been the most widely used diagnostic technique for pulmonary and cardiac disorders in healthcare systems around the world, owing to its low cost and ease of use. However, the accuracy of the diagnosis is severely hampered by a lack of radiologists and inter-observer variability, which is exacerbated by resource-constrained circumstances. Even though deep learning methods for disease classification have shown great promise, there has been a lack of rigorous comparative evaluations of contemporary architectures for prediction using balanced multi-class chest disease data.

**Aims**: This work examines seven popular deep learning models for multi-class Chest X-ray classification, with an emphasis on trade-offs between performance indicators and computational efficiency. The findings would inform deployment decisions in healthcare settings with diverse resource availability.

**Methodology**: The architectures studied were ConvNeXt-Tiny, DenseNet121, DenseNet201, ResNet50, Vision Transformer (ViT-B/16), EfficientNetV2-M, and MobileNetV2. A comprehensive dataset was generated from existing repositories, consisting of 13,108 training photos, 1,455 validation images, and 3,517 test images for five conditions: Cardiomegaly, COVID-19, Normal, Pneumonia, and Tuberculosis. All models were initialized with ImageNet-pretrained weights and trained under consistent settings, including standardized preprocessing, data augmentation, and optimization hyperparameters. To evaluate model performance, we used metrics such as AUROC, overall accuracy, precision, recall, F1-score, and computational efficiency.

**Results**: All seven studied designs achieved test accuracies that exceeded 90%. ConvNeXt Tiny performed well, with a validation AUROC of 98.64% and a test accuracy of 92.31%. ResNet50 followed closely with 92% test accuracy, while ViT-B/16 earned 91.87%. Notably, MobileNetV2 emerged as the most parameter-efficient Net alternative, with only 3.50 million parameters. Despite its small size, it achieved a test AUROC of 94.10% and obtained the highest efficiency rating in our study. This lightweight architecture achieved around 98.3% of the accuracy of the best-performing model while using 87.5% fewer parameters, which has important implications for deployment in resource-constrained contexts.

**Conclusion**: The current findings show that excellent accuracy in multi-disease categorization of Chest X-ray pictures is possible without requiring significant computational resources. This finding has important implications for the practical integration of deep learning as a diagnostic aid in a variety of healthcare settings, both resource-rich and resource-constrained. Furthermore, choosing a suitable architecture should consider the available infrastructure as well as the unique characteristics of each deployment scenario.

## 1. Introduction

One of the health problems, which ranks highest in the world, is chest disease, where hundreds of millions of people are diagnosed every year, and they carry significant social and economic implications. The diseases are high-morbidity and high-mortality illnesses, including respiratory and heart infections. Some of the most important ones are Pneumonia, Tuberculosis (TB), COVID-19, and Cardiomegaly due to the number of deaths they cause. The WHO estimates a world death toll of about 2.5 million annually as a result of Pneumonia, with nearly 700,000 of these being children below the age of five years [1]. TB is among the most serious global crises, where 10.6 million infected individuals are diagnosed, and 1.3 million deaths are reported in 2022, in South-East Asia and Africa [2]. Despite being relatively new, COVID-19 has already claimed the lives of millions of people and is placing a burden on the healthcare systems of the world [3]. As a little-known fact among the population, Cardiomegaly usually presupposes cardiac failure and early death [4]. All of these statistics suggest the need to introduce high-quality, affordable, and comprehensive diagnostic methods for chest diseases.

Radiography is not the only diagnostic pathway that can be used to diagnose Pneumonia; other non-radiographic diagnostic methods include cross-sectional imaging, which can be Computed Tomography (CT) and Magnetic Resonance Imaging (MRI), infectious causes, which may involve sputum culture and blood analysis, and cardiac assessment, which may include echocardiography [5]. The first-line diagnostic method is Chest X-ray (CXR), which is readily available and cost-effective, particularly in facilities with limited resources [6]. However, the manual subjective interpretation can be erroneous. Radiographic appearances often overlap with disease appearances, e.g., patchy opacities in TB and Pneumonia; other factors, such as the radiographer's experience, workload, and fatigue, also contribute to inter-observer variation [7], [8]. The consequences of such restrictions include delayed or missed diagnoses, especially in high-volume clinical settings. The importance of this study is that it addresses critical issues in diagnosis by leveraging advanced Artificial Intelligence (AI) techniques that deliver consistent, rapid, and precise disease classification, especially in healthcare settings with scarce resources and limited diagnostic knowledge.

Over the last decade, medical image recognition has seen a revolution as AI, and more precisely, Deep Learning (DL), has replaced manual interpretation by radiologists. Convolutional Neural Networks (CNNs) and hybrid networks can also detect finer image details that the naked eye cannot and evaluate a radiologist's performance in specialized diagnostics [9], [10] [11]. CheXNet (DenseNet-121) is notable for achieving high accuracy on the detection of Pneumonia in the NIH ChestX-ray14 dataset, leading to the subsequent research on AI-assisted interpretation of CXRs [12]. Later models, such as EfficientNet and CNN-Transformer hybrids, were used to improve feature representations and generalization [13], [14]. Their existence notwithstanding, the multi-class classification of chest diseases often results in reduced performance due to feature overlap, class similarity, and class imbalance, underscoring the need for balanced datasets to ensure fair and accurate diagnoses, which is crucial for healthcare professionals and AI developers [15]. Only a few studies directly compared modern CNN and Transformer architectures under the same training conditions on a balanced multi-class CXR dataset.

A principal challenge is constructing robust datasets for categorizing chest images. Some publicly available datasets include NIH ChestX-ray14 [16] COVID-19 Radiography Database [17], and Mendeley Tuberculosis Database [18] They differ in terms of image quality, label consistency, and classification. Typical cases are frequently overrepresented, whereas Cardiomegaly and Tuberculosis are underrepresented, resulting in a dataset imbalance that can enable models to overfit on majority classes while underperforming on minority classes, raising concerns about clinical dependability [19], [20]. Such skewed data can bias models to produce deceptively high accuracy but low recall for minority classes. To address this, our study uses measurements such as F1-score and recall, which more accurately reflect model performance across all classes, particularly underrepresented disorders, assuring therapeutic relevance [21], [22]. Variations in imaging devices, exposure conditions, and annotation standards between datasets limit cross-domain generalization and increase complexity in the actual world for practical applications. We use cross-dataset validation and domain adaptation strategies to improve model robustness and clinical applicability [23]. As a result, preprocessing, balancing, data growth, and robust optimization are critical for creating dependable AI systems for CXR analysis [24], [25].

This study makes the following key contributions:

- We construct a balanced and unified five-class Chest X-ray dataset by integrating multiple public repositories, effectively addressing class imbalance while preserving the diversity representative of real clinical practice.
- The study systematically evaluates state-of-the-art CNN and transformer-based models to identify the most clinically viable approach for multi-disease Chest X-ray classification.
- The CheXNet protocol is reformulated from a multi-label to a single-label disease classification framework, improving interpretability and clinical applicability.

The rest of this paper is organized as follows. Section 2 reviews the related literature on AI-based Chest X-ray classification. Section 3 describes the dataset, preprocessing steps, and the proposed framework. Section 4 presents and discusses the experimental results. Finally, Section 5 concludes the paper and outlines future perspectives.

## 2. Related Work

Deep learning in the analysis of chest radiographs has advanced significantly in the last few years. The studies in this field have taken several directions simultaneously, such as architectural advancement, improved computational efficiency, and the development of multi-disease classification features. The initial studies determined that CNNs would be as effective as radiologists practicing on each specific diagnostic task with suitably large datasets [26]. This initial success triggered extensive further research, but most of the work done has focused on testing a single architecture or comparing just a few models under different experimental conditions. As a result, the systematic evaluation of architectural trade-offs and their clinical implementation has not yet been thoroughly carried out.

More recent attempts have sought to replicate and refine these basic methods [27]. Much valuable information about the consistency and generalizability of the previous results has been obtained from these reproduction studies. However, they have also demonstrated the difficulties of achieving consistent performance across datasets and preprocessing protocols.

### 2.1. Modern CNN Architecture and EfficientNet Networks

The architecture of CNNs has undergone several changes to improve accuracy and efficiency. EfficientNet proposed a principled way to scale up a model: it optimizes the model architecture by balancing network depth, width, and resolution using compound coefficient optimization [28]. This compound-scaling approach showed that coordinated scaling across 3 dimensions yields higher performance than arbitrary scaling of each aspect. EfficientNet applications in medical imaging, especially Chest X-ray classification, have demonstrated strong competitiveness while requiring significantly lower computational resources than previous architectures [29], so they are instrumental in resource-constrained clinical settings.

ConvNeXt is a modern architecture aiming to update the standard convolutional architectures with design principles of Vision Transformers without sacrificing the efficiency and simplicity of CNNs [30]. Through systematic study and adaptation of desirable features of transformer architectures, such as adapted training regimes, macro design motifs, and inverted bottleneck architectures, ConvNeXt has shown that pure convolutional networks can compete with transformer-based architectures on several computer vision tasks. This architecture bridges the gap between classical CNNs and contemporary transformers, offering an interesting alternative for medical image analysis, where both precision and computational efficiency matter.

### 2.2. Vision Transformers in Medical Imaging

With the introduction of Vision Transformers (ViT), attention-based processes have emerged as a viable alternative to the standard convolutional paradigm for image analysis [31]. Unlike CNNs, which extract local features in a hierarchical sequence, transformers utilize self-attention to construct long-range dependencies throughout the entire image domain, potentially exposing subtle diagnostic patterns across several anatomical locations. Recent investigations have thoroughly investigated the usage of transformers in medical imaging [32], demonstrating their performance in a variety of diagnostic tasks, including segmentation, classification, and detection. However, these studies highlight substantial trade-offs: transformers frequently require larger training corpora than CNNs to achieve comparable performance, as well as greater processing resources for both training and inference.

Comparisons of CNNs and transformers in medical image classification have revealed subtle performance differences [33]. Transformers are also excellent encoders of global context and long-range spatial relationships, which help detect diffuse pathological patterns in chest radiographs. Still, they can fail on small medical image datasets, where CNNs' inductive biases are more useful. The results highlight that architectural choices must not rely solely on accuracy measures but also on data accessibility, computing capabilities, and diagnostic needs.

### 2.3. Multi-Class Classification of Chest Diseases

Multi-class classification of chest diseases is not as straightforward as in binary disease detection or multi-label classification. Another lightweight six-layer convolutional neural network specifically trained to classify six thoracic conditions was presented in [33], achieving 80% accuracy on multi-class tasks and 97.9% on binary Pneumonia detection. This paper has shown that task-reconfigurable lightweight architectures can be clinically acceptable and consume significantly less computational resources than more complex networks. In a parallel study, a variety of architectures, including VGG16, InceptionResNetV2, and a home-built CNN, were tested to identify differences between COVID-19, Pneumonia, and Normal cases [34]. The custom network achieved 97% accuracy and 98.21% reported sensitivity to the evaluation set. In [35], the study achieved a validation accuracy of 98.72%. Although the performance figures in these studies were impressive in their respective tasks, they tested only one or two architectures each, which provides little insight into the performance of different design philosophies and architectural approaches under similar experimental conditions.

### 2.4. Hybrid and Ensemble Approaches

Hybrid and ensemble strategies have been studied to combine multiple learning mechanisms or model types to leverage complementary strengths. Nair and Singh were the first to use CNN-LSTM to combine spatial and temporal features [36], achieving 99% recall and 94% F1-score on six thoracic diseases. Likewise, Sharma and Kamble integrated CNN with LSTM models [37], obtaining an accuracy of 98.5% with multi-class Chest X-ray classification and demonstrating the ability of hybrid structures to both learn spatial and temporal information of medical images. A different approach and combined transfer learning were used with generative adversarial networks to address data scarcity in COVID-19 detection [38], [39], with the accuracy of 98% and successful data augmentation. Ensemble learning is used to combine handcrafted features with deep learning classifiers [40], which achieved 98% accuracy at a lower computational cost than deep learning alone. Knowledge distillation is used to build small student networks [41], with an accuracy of 96.08% and a significantly smaller model size and lower computation. The fusion architectures examined were CheXNet, which was combined with feature pyramid networks to classify multi-label Chest X-rays [42], and explained the benefits of combining several architectural structures and designs. These hybrid designs illustrate how to trade off diagnostic performance for computational efficiency, but the complexity may be impractical for clinical practice.

### 2.5. Interpretability and Uncertainty Quantification

The concepts of interpretability and uncertainty quantification have increasingly gained importance in recent scholarship as predictive models are increasingly used in clinical settings. A graph-based framework for modeling spatial-semantic relationships between anatomical regions in the MIMIC-CXR dataset was illustrated, achieving a mean area under the receiver operating characteristic curve (AUC) of 86.09% and providing specific uncertainty estimates [43]. This is a significant change from accuracy-only optimization; however, such models usually have higher computational costs and require more detailed implementation.

### 2.6. Disease-Specific Systems

Specific disease diagnostic systems have shown high performance in well-defined pathological areas. A MobileNet-based Tuberculosis detection pipeline was created and had a range of AUC values between 95.1% and 97.5% [44]. A systematic review of 54 studies on deep learning in Tuberculosis screening was conducted, finding that current CNN models can achieve cumulative AUC scores of up to 99% on self-contained benchmark datasets. Though these specialized systems are good at diagnosing specific target pathologies, they are not as general as needed for multi-disease screening [45].

### 2.7. Research Gaps and Study Contributions

These are key methodological gaps highlighted by the current literature review. First, most studies evaluate a single or two architectures, which prevents them from making meaningful comparisons between divergent design philosophies, e.g., classical convolutional networks and modern transformers, under the same experimental conditions. Second, training programs, preprocessing pipelines, and data-augmentation techniques differ significantly across research, making it challenging to directly compare performance. Third, most studies use vastly imbalanced datasets where Normal samples far outnumber pathological ones; this bias can overstate reported accuracy rates as well as obscure poor performance on minority disease categories. Fourth, the systematic study of the trade-offs among model complexity, computational requirements, and diagnostic fidelity remains limited. Moreover, it is conceptually relevant to the deployment choices for heterogeneous healthcare infrastructures. All these shortcomings underscore the need for regulated, comparative experiments with standardized procedures.

To overcome these drawbacks, we have conducted a systematic comparison of seven modern neural architecture designs that represent different design philosophies. The selected architectures are dense connectivity patterns of DenseNet121 and DenseNet201, residual learning of ResNet50, computational learning of MobileNetV2, optimized scaling of EfficientNet V2M, modern convolutional learning of ConvNeXt Tiny, and attention-based learning of Vision Transformer ViT-B/16. Each of the seven models was trained under strictly standardized conditions, using the same preprocessing algorithms, augmentation policies, and optimization hyperparameters on a highly balanced dataset comprising five disease subtypes: Cardiomegaly, COVID-19, Normal, Pneumonia, and Tuberculosis. This experimental design enables a straightforward assessment of the architectural virtues and shortcomings, as well as the measurement of performance-efficiency trade-offs relevant to clinical implementation across a wide variety of resource conditions. Table 1 presents the main features of recent research in the field and shows how our multi-architectural comparison, implemented using the same protocols, directly addresses the gaps identified in the study.

**Table 1:** Comparison of Recent Deep Learning Approaches for Chest X-ray Classification

| Ref | Year | Approach | Main Gap Addressed | Disease Classes | Best Reported Metric Value (%) |
|---|---|---|---|---|---|
| [26] | 2018 | DenseNet-121, multi-label classification | Baseline architectural evaluation | 14 thoracic diseases | AUC 84-88 |
| [45] | 2022 | Systematic review of TB systems | Single disease focus | TB only | Up to 99 |
| [41] | 2023 | Lightweight 6-layer CNN | Single architecture only | 6 classes | Acc 80 |
| [35] | 2023 | Unified multi-disease CNN | Single architecture only | 4 classes | Acc 98.72 |
| [37] | 2023 | CNN-LSTM hybrid architecture | Limited architectural diversity | Multiple classes | Acc 98.5 |
| [43] | 2023 | Graph convolutional network | High complexity, limited scalability | 14 classes | AUC 86.09 |
| [34] | 2024 | VGG16, Inception, custom CNN | Limited comparison (3 models only) | 3 classes | Acc 97 |
| [40] | 2024 | Ensemble with handcrafted features | Not end-to-end deep learning | 3 classes | Acc 98 |
| [38] | 2024 | Knowledge distillation | Single architecture type only | 4 classes | Acc 96.08 |
| [42] | 2024 | CheXNet with a feature pyramid network | Single fusion architecture only | Multilabel | AUC 84.6, Accuracy 91.4 |

| [44] | 2024 | MobileNet for Tuberculosis | Single disease focus | TB vs Normal | AUC 95.1-97.5 |

## 3. Materials and Methods

### 3.1. Dataset Description

A five-class Chest X-ray dataset was constructed using three publicly available repositories, including Normal cases and four thoracic pathologies: Pneumonia, Tuberculosis, COVID-19, and Cardiomegaly. From the NIH Clinical Center Chest X-ray dataset, which provides multi-label annotations, only frontal-view images annotated with a single target label (Normal, Pneumonia, or Cardiomegaly) were selected. Images containing multiple findings were excluded to ensure label consistency for single-label multi-class classification. COVID-19 images were obtained from the COVID-19 Radiography Database, and Tuberculosis images were sourced from the Tuberculosis Chest X-ray Database to complete the remaining classes. Owing to differences in dataset size and class prevalence among the repositories, the resulting merged dataset exhibited class imbalance, as detailed in Table 2.

- NIH Clinical Center Chest X-ray Dataset [16]This large dataset of frontal-view Chest X-ray images of 30,805 unique patients (112,120 photos) included images classified as Cardiomegaly, Normal, and Pneumonia.
- COVID-19 Radiography Database [17]The curated collection of Covid-19 Chest X-rays used to obtain the images was Covid-19 radiographs, which were annotated to identify Covid-19.
- Tuberculosis Chest X-ray Database [18]This specific database was used to access cases of Tuberculosis through a routine and TB-positive chest radiograph from various institutions.

This combination of heterogeneous sources created an initial dataset that had high class imbalance, as discussed in Table 2.

Table 2: Original dataset class distribution before balancing

| Class Name | Number of Images | Resources |
|---|---|---|
| Normal | 60,361 | [16] |
| Pneumonia | 1,390 | [16] |
| Cardiomegaly | 2,735 | [16] |
| COVID-19 | 3,616 | [17] |
| Tuberculosis | 2,494 | [18] |

### 3.2. Data Preprocessing Steps

The dataset underwent several preparatory steps to improve its quality, balance, and support machine learning. These steps are described in the following sections:

### 3.2.1. Random Undersampling of the Majority Class:

To mitigate the severe class imbalance present in the original dataset (60,361 Normal images versus 1,390 Pneumonia images), random undersampling was applied to the Normal class. Prior studies on CNN training have shown that extreme class imbalance can bias the learned decision boundaries toward majority classes, resulting in inflated overall accuracy but poor sensitivity to minority categories [21], [46].

To explicitly counter this effect, random undersampling without replacement was applied to the Normal class, reducing it to 3,616 images to match the size of the largest minority class (COVID-19). This choice was made to ensure a balanced optimization landscape across all classes, rather than allowing the model to overfit redundant Normal samples. Notably, the reduced sub-set size remains statistically adequate for transfer learning, where pretrained feature extractors substantially reduce the dependence on large class-specific sample sizes.

This balancing strategy promotes a more equitable learning process, reduces redundancy in the majority-class samples, and improves performance metrics such as recall and F1-score across all disease categories, thereby enhancing the clinical reliability of the model.

### 3.2.2. Oversampling via Controlled Data Augmentation

To address class imbalance among the minority categories (Pneumonia, Tuberculosis, COVID-19, and Cardiomegaly), a controlled oversampling strategy using data augmentation was employed. All original images in each minority class (n ≤ 3,616) were preserved to ensure genuine pathological characteristics were retained and not distorted by synthetic data.

An augmentation pipeline was implemented using Torchvision Transforms, incorporating random horizontal flipping (probability = 0.5), random rotations within ±15°, and brightness and contrast adjustments limited to ±10% [44]. These transformations were constrained to reflect variations commonly observed in real-world clinical settings.

Augmentation was applied iteratively until each minority class reached 3,616 images, aligning the class distributions with the undersampled Normal class. This low-intensity augmentation strategy was selected to enhance minority-class representation while minimizing the risk of overfitting to artificially generated patterns. By maintaining a high proportion of original images and limiting transformation magnitude, the augmented samples preserved clinical plausibility and diagnostic relevance, as illustrated in Figure 1.

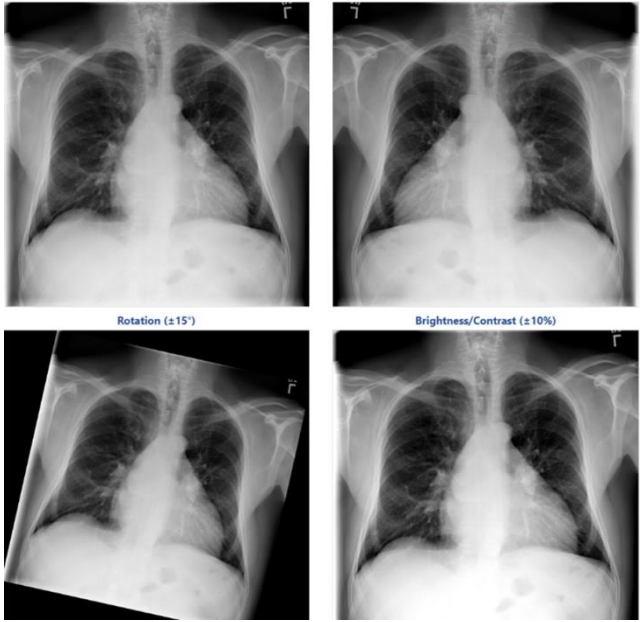

**Figure 1:** Oversampling strategy using controlled data augmentation applied to minority classes.

Table 3 illustrates the distribution of our Chest X-ray dataset before and after balancing. First, the class balance was extreme, with the Normal class having a much larger sample size than the other groups. To obtain a balanced dataset suitable for machine learning, we used random undersampling to equalize the Normal class with the largest minority class, and systematic controlled augmentation to increase representation of underrepresented classes. Consequently, all disease classes now have an equal number of samples, providing a fairer and more valid basis for future model development and testing.

**Table 3:** Chest X-ray dataset class distribution before and after balancing.

| Class Name | Before Balancing | After Balancing |
|---|---|---|
| Normal | 60361 | 3616 |
| Pneumonia | 1390 | 3616 |
| Covid-19 | 3616 | 3616 |
| Tuberculosis | 2494 | 3616 |
| Cardiomegaly | 2735 | 3616 |

### 3.2.3. Dataset Partitioning and Patient Isolation

To prevent data leakage and ensure unbiased evaluation, a strict patient-level data splitting strategy was adopted. Unique patient identifiers were extracted from image filenames, yielding 18,080 images from 6,978 patients (≈2.6 images per patient). The dataset was partitioned at the patient level into 70% for Training, 10% for Validation, and 20% for Testing, ensuring complete patient isolation across splits (train ∩ test = ∅, validation ∩ test = ∅). This resulted in 5,582 patients for training and validation and 1,396 patients for testing, corresponding to 14,563 and 3,517 images, respectively.

Following class balancing (3,616 images per class), the same patient-level split was preserved. The training set was used to learn model parameters, the validation set for hyperparameter tuning and overfitting assessment, and the test set—fully isolated from model development—for final performance evaluation. This protocol reflects real clinical deployment, where models are applied to previously unseen patients, and provides a reliable estimate of generalization performance (Table 4, Figure 2).

Rather than k-fold cross-validation, we adopted a strict patient-level dataset partitioning strategy. The dataset was divided into 70% Training, 10% Validation, and 20% Testing at the patient level, with complete patient isolation in the test set (train ∩ test = ∅). This design prevents patient-level data leakage and reflects real clinical deployment, where models encounter previously unseen patients. The dataset size (18,080 images from 6,978 patients) provides sufficient data for stable performance estimation under this split.

**Table 4:** Dataset Split Statistics with Patient-Level Isolation

| Split | Images | % of Total | Unique Patients | % of Patients |
|---|---|---|---|---|
| Training Validation | 14,563 | 80 | 5582 | 80 |
| Testing | 3,517 | 20 | 1,396 | 20 |
| Total | 18,080 | 100 | 6,978 | 100 |

Having completed the class-balancing process, in which each disease category had a total of 3,616 images, the dataset was further split into three subsets: Training (≈ 70%), Validation (≈ 10%), and Testing (≈ 20%), as shown in Table 4. This separation adheres to existing deep-learning standards for medical imaging, ensuring that sufficient data has been collected to optimize the model while maintaining a separate set for independent evaluation [47], [48].

Learning the network parameters and capturing disease-specific visual patterns was done using the training set, which includes most of the data. The validation set was used during model development to optimize hyperparameters, assess convergence, and evaluate overfitting by applying the model to unseen data. The test set was not used for training and was only evaluated at the end of the model, yielding a reasonable estimate of generalization performance [49].

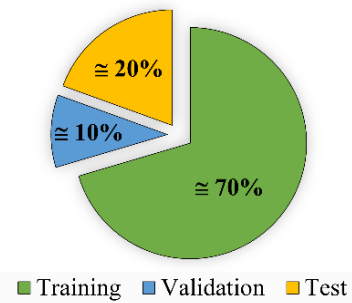

**Figure 2:** Dataset split distribution using a 70% Training, 10% Validation, and 20% Testing partition at the patient level.

Figure 2 shows that the 70:10:20 split is a viable balance between learning efficiency and evaluation reliability. The larger the training part, the more robust the optimization; the more time is allocated to validation and testing, the lower the computational cost, and the less information leakage between the subsets. Also, stratified sampling was used to maintain equal representation of each class across subsets, a necessary measure to avoid bias in multi-class medical data [50].

### 3.3. Comparative Framework Overview

This paper presents a controlled comparative study of seven deep learning networks for classifying Chest X-rays into multiple diseases. We modified the original CheXNet system to a single-label classification paradigm rather than a multi-label one, enabling us to make specific diagnoses of five diseases: Cardiomegaly, COVID-19, Normal, Pneumonia, and Tuberculosis. All models were trained and assessed under the same conditions to provide a fair comparison using a single preprocessing pipeline, a standardized training regime, and standard evaluation measures. Seven state-of-the-art architectures were compared, which are of various methods of extracting image features:

#### 3.3.1. Model Pipeline

The algorithm will be divided into collecting Chest X-ray data from the public, equalizing across five disease groups, and splitting the data into training and test subsets. Several deep learning models are then fine-tuned and tested using the same steps, and each model's performance is evaluated against key metrics to compare its ability to detect chest diseases. Furthermore, the total workflow (shown in Figure 3) consists of a training pipeline that depicts the iterative optimization process and a testing pipeline that evaluates the models' performance using multi-crop inference, enabling complete, consistent evaluation.

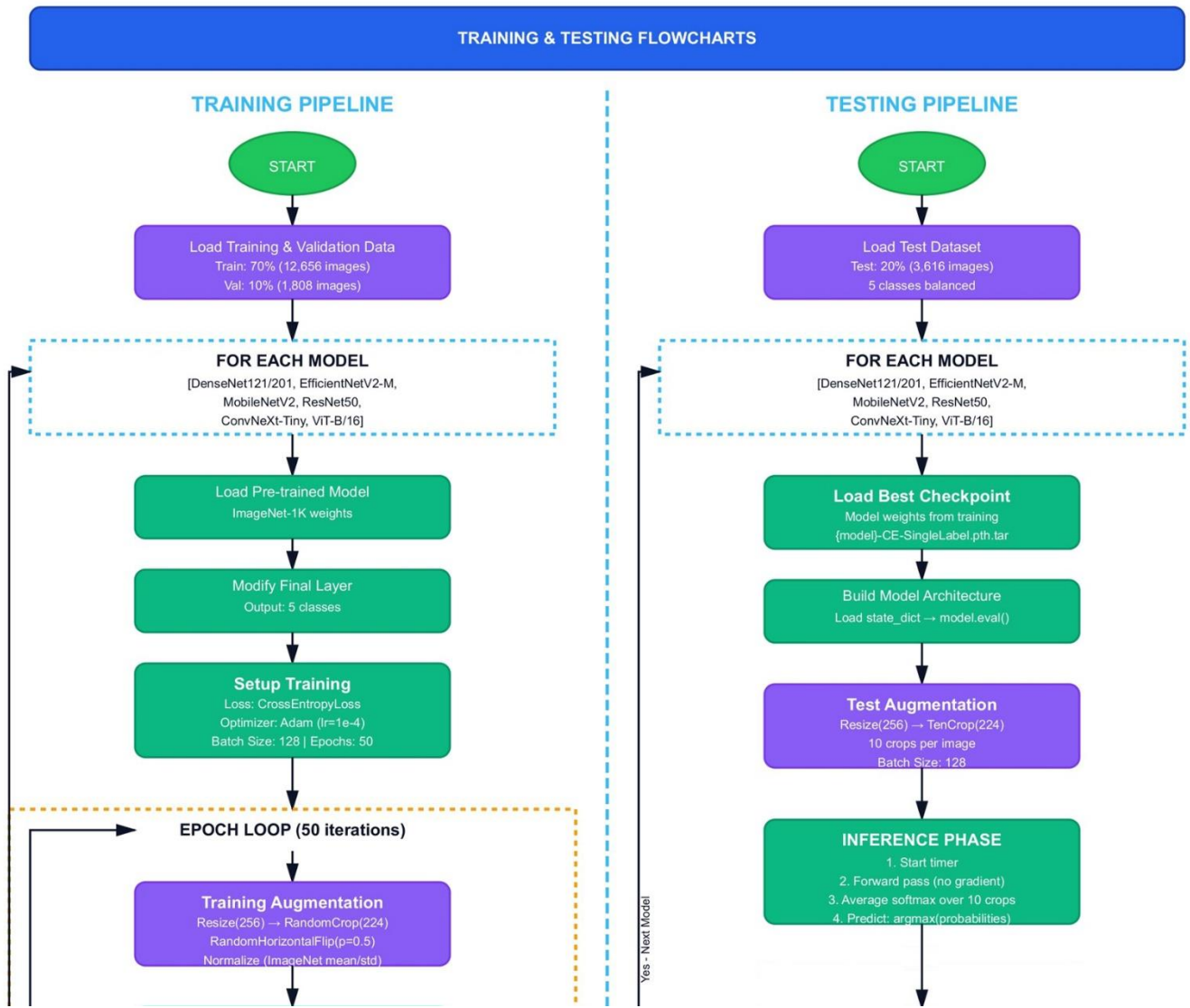

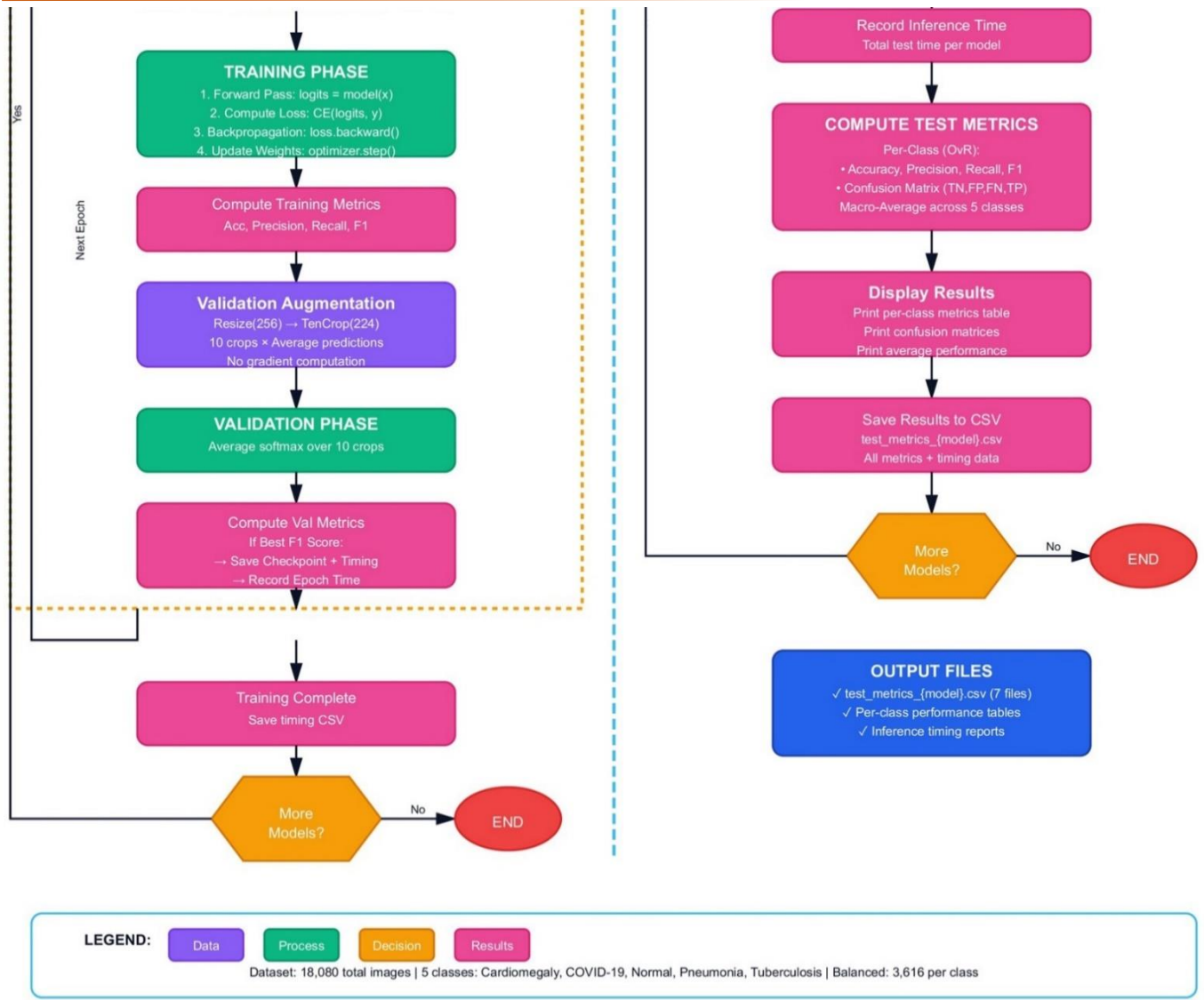

**Figure 3:** Complete training and testing pipeline for multi-disease chest X-ray classification

### 3.3.2. Modified CheXNet-Based Backbone Architecture

The proposed architecture is derived from the original CheXNet framework and adapted to support single-label, five-class Chest X-ray classification, as illustrated in Figure 4. While CheXNet was initially designed as a multi-label classification model, we modified its architecture to operate in a unified single-label setting suitable for comparative evaluation across different backbones.

A standardized preprocessing pipeline is applied to all input images, after which features are extracted using one of seven backbone networks: DenseNet121, DenseNet201, EfficientNetV2-M, MobileNetV2, ResNet50, ConvNeXt-Tiny, or ViT-B/16. In each case, the original CheXNet classification head is replaced with a task-specific fully connected layer producing five logits corresponding to the target disease categories.

During inference, TenCrop evaluation is employed, and the predicted probabilities across crops are averaged to enhance prediction stability and robustness. This modified CheXNet-based framework allows all backbone architectures to be evaluated under identical training and inference conditions, ensuring a fair and clinically relevant comparison.

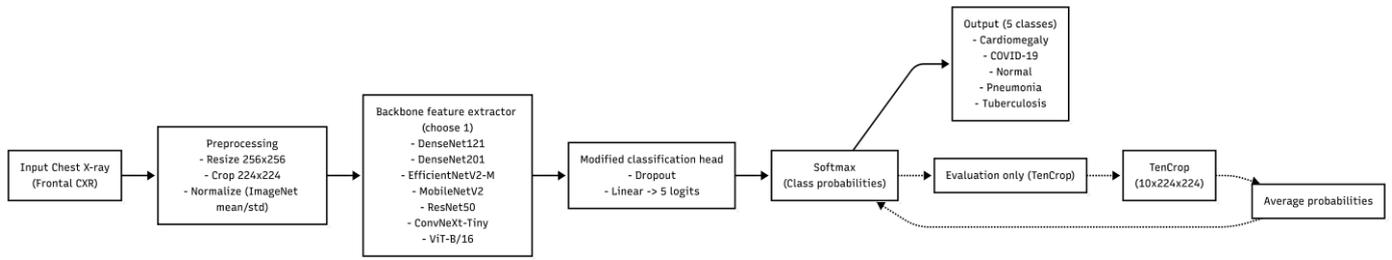

**Figure 4:** Proposed CheXNet-based backbone replacement framework for five-class chest X-ray classification.

### 3.3.3. DenseNet Convolutional Networks

- **DenseNet121**: The original CheXNet backbone [12], which has 121 layers with dense connectivity patterns in which each layer takes in feature maps of all the previous layers [51]. This architecture is the one we compare against.
- **DenseNet201**: DenseNet201 is a more intensive version of DenseNet featuring 201 layers, which yet again has more effective feature extraction ability due to the addition of more dense blocks, and retains the same connectivity pattern as DenseNet121 [52].

### 3.3.4. EfficientNet and MobileNet Architectures

- EfficientNetV2-M: A medium-scale EfficientNet family model that uses compound scaling to trade off network depth, width, and resolution [28]. The new system, EfficientNetV2, proposes training-aware neural architecture search and progressive learning, which improve training efficiency and accuracy over prior methods [53].
- MobileNetV2: A lean Mobile and resource-constrained architecture. It employs inverted residual blocks, which are linear bottlenecks, and depthwise separable convolutions to reduce computational time while achieving competitive accuracy [54].

### 3.3.5. Residual Networks

- **ResNet50:** A 50-layer residual network that employs skip connections to address the vanishing gradient problem in deep networks. ResNet-50 is a well-performing architecture widely used as a baseline in a wide range of computer vision tasks [10].

### 3.3.6. Modern CNN and Transformer Architectures

- ConvNeXt-Tiny: A fully convolutional network that is modernized and uses the design ideas of Vision Transformers, with the efficiency of CNNs [30]. ConvNeXt employs larger kernel sizes, inverted bottleneck structures, and fewer activation functions, achieving Transformer-competitive performance with standard convolutional operations.
- ViT-B/16: Vision transformer with base configuration, which splits input images into 16x16 patches and processes them with self-attention mechanisms [31]. ViT is a paradigm shift in the approach to pure convolution, using attention-based feature extraction to model global context at the first layer.

### 3.4. Model Adaptation and Configuration

All architectures were fine-tuned for our five-class Chest X-ray classification task, using models pre-trained on ImageNet-1K. Transfer learning from ImageNet has been shown to substantially improve performance on medical imaging tasks, even with limited labeled data.

Each model was customized by modifying its classifier to output predictions for the five disease categories:

- DenseNet121/201: The original single fully connected layer was replaced with nn.Linear(in_features, 5).
- EfficientNetV2-M: The classifier head was reconstructed as nn.Sequential(nn.Dropout(p=0.0), nn.Linear(in_features, 5)).
- MobileNetV2: The classifier retained its dropout regularization: nn.Sequential(nn.Dropout(p=0.2), nn.Linear(in_features, 5)).
- ResNet50: The final fully connected layer (fc) was replaced with nn.Linear(2048, 5).
- ConvNeXt-Tiny: The three-layer classifier head was adapted with the final linear layer modified to output five classes.
- ViT-B/16: The transformer head was modified to nn.Linear(768, 5), converting patch-level representations into disease predictions.

All models were implemented using PyTorch's torchvision.models library with official pre-trained weights (IMAGENET1K_V1) to leverage transfer learning. Models were parallelized across multiple GPUs using torch.nn.DataParallel to accelerate training and inference. Each architecture was trained separately following the same dataset and training procedure.

### 3.4.1. Hyperparameters and Training Configuration

The model was trained using the Adam optimizer with the configuration summarized in Table 5.

**Table 5:** Summary of Model Training Parameters

| Component | Description |
|---|---|
| Loss Function | CrossEntropyLoss for single-label classification |
| Optimizer | Adam optimizer |
| Adam Parameters | $\beta_1 = 0.9$, $\beta_2 = 0.999$, $\varepsilon = 1e-8$ |
| Learning Rate | 1e-4 (fixed throughout training) |
| Weight Decay | 1e-5 (L2 regularization) |
| Batch Size | 128 (training), 128 (testing) |
| Epochs | 50 |
| Early Stopping Criterion | Best model selected based on the highest validation AUROC; ties resolved using macro-averaged F1-score, then accuracy. |

Adam optimizer was selected for its adaptive learning rate and its effectiveness in classifying medical images. The fixed learning rate is 1e-4, chosen according to standard practice in transfer learning scenarios when pre-trained weights must be fine-tuned [47], [55].

### 3.4.2. Data Loading and Preprocessing Configuration
Table 6 Data loader and preprocessing pipeline hyperparameters.

**Table 6:** Data loader and preprocessing pipeline hyperparameters.

| Component | Description |
|---|---|
| Number of Workers | 8 parallel processes for data loading |
| Pin Memory | Enabled for faster GPU transfer [56] |
| Persistent Workers | Enabled to maintain worker processes across epochs |
| Training Shuffle | Enabled to prevent ordering bias [57] |
| Validation/Test Shuffle | Disabled to ensure reproducible evaluation [50] |

### 3.4.3. Evaluation Metrics and Analysis
Model performance was evaluated using comprehensive metrics computed in a one-vs-rest (OvR) framework for each disease class [58]:

Per-class metrics:

- Accuracy measures the proportion of correctly predicted images out of all predictions [59], and calculated as:

$$Accuracy = \frac{(TN+TP)}{(TN+FP+FN+TP)} \quad (1)$$

Where:
TP = True Positives
TN = True Negatives
FP = False Positives
FN = False Negatives

- Precision is used to determine how accurately a medical image classification model predicts the actual positive image in all the photos that the model predicts to be positive. It indicates the proportion of correct predictions of the model [60].

$$Precision = \frac{TP}{(TP+FP)} \quad (2)$$

- Recall (Sensitivity) in medical image classification is used to determine the percentage of real positive cases (diseased images) that the model can detect properly. It demonstrates the extent to which the model explains all the actual cases of the condition[61].

$$Recall\ (Sensitivity) = \frac{TP}{TP+FN} \quad (3)$$

- The harmonic mean of precision and recall is the F1-Score in medical image classification, which gives one metric that balances the other two metrics. It can particularly be handy in case of imbalanced classes, since it punishes extreme values of either precision or recall [59].

$$F1 - Score = 2 \times \frac{Precision \times Recall}{Precision + Recall} \quad (4)$$

Aggregate metrics:

- Macro-averaged accuracy, precision, recall, and F1-score: Unweighted mean across all five classes, treating each disease equally regardless of sample size [62].

### 3.5. Implementation Details
Models were trained and tested on the setup in Table 7 using Google Colab GPUs and PyTorch, with consistent splits, preprocessing, and hyperparameters. Performance metrics and best model checkpoints were logged to ensure reproducibility and enable fair comparison in multi-disease Chest X-ray classification.

**Table 7:** Computational setup, data handling, and model management details.

| Category | Details |
|---|---|
| Software Environment | Framework: PyTorch 2.x with torchvision library |
| | Development platform: Google Colab with GPU acceleration |
| | CUDA enabled, cudnn.benchmark=True |
| Hardware Specifications | Google Colab GPU instances (NVIDIA Tesla T4/V100) |
| | Batch processing parallelized via DataParallel. |
| Reproducibility Measures | Data split: 70% train, 10% validation, 20% test. |
| | Identical splits across all models |
| | Standardized preprocessing & augmentation |
| | Fixed hyperparameters |
| Performance Tracking | Per-epoch training time recorded (seconds) |
| | Cumulative training time tracked |
| | Inference time measured for the complete test set |
| | All timing data are saved in CSV format. |
| Model Checkpointing | Naming: {backbone}-AccPrecRecF1-ConfMat-CE-SingleLabel_2026.pth.tar |
| | Contains: weights, best validation accuracy/F1, epoch, timing, class details |

This methodological approach has ensured that the seven architectures are tested using pure independent tests,

enabling fair and reasonable performance comparisons for multi-disease Chest X-ray classification.

## 4. Results and Discussions

### 4.1. Results

This section presents an in-depth comparison of seven state-of-the-art deep learning architectures for the classification of various diseases in Chest X-rays. It was assessed on a multi-source, equal-weight dataset, and all models were subjected to standardized experimental procedures to ensure fair comparisons. The training and validation

Table 8 summarizes the performance of all seven architectures on the validation dataset. All models achieved high accuracies above 94.9%, demonstrating the effectiveness of the balanced multi-source dataset and the consistent training pipeline. The experimental framework's robustness is datasets contained 14,563 samples, and the test dataset contained 3,517 samples across five disease types: Cardiomegaly, COVID-19, Normal, Pneumonia, and Tuberculosis. These architectures were trained under equal preprocessing pipelines, a common optimization strategy, that is, the Adam optimizer with a learning rate of 0.0001, and equal evaluation metrics, and therefore, confounding variables were reduced to a minimum, and a direct architectural comparison was possible.

#### 4.1.1. Overall Performance Analysis

further supported by strong performance across multiple architectures, including traditional CNNs (ResNet50, DenseNet), EfficientNet models (MobileNetV2, EfficientNetV2-M), modern CNNs (ConvNeXt-Tiny), and vision transformers (ViT-B/16).

**Table 8:** Overall Performance Metrics on Validation Dataset

| Model | Parameters (Million) | Validation. Accuracy (%) | Validation. Precision (%) | Validation. Recall (%) | Validation. F1-Score (%) | Validation. AUROC (%) | Epochs | Training Time |
|---|---|---|---|---|---|---|---|---|
| ConvNeXt-Tiny | 27.82 | 96.25 | 90.28 | 89.95 | 90.54 | 98.64 | 50 | 1h 43m |
| DenseNet121 | 6.96 | 95.92 | 90.07 | 89.84 | 89.78 | 98.43 | 50 | 1h 36m |
| DenseNet201 | 18.32 | 95.89 | 89.94 | 89.78 | 89.61 | 98.49 | 50 | 2h 48m |
| ResNet50 | 23.52 | 95.57 | 89.20 | 89.23 | 88.88 | 98.33 | 50 | 1h 11m |
| EfficientNetV2-M | 52.86 | 95.13 | 88.26 | 88.20 | 88.04 | 97.98 | 50 | 2h 52m |
| ViT-B/16 | 85.80 | 94.95 | 88.70 | 87.25 | 87.58 | 98.06 | 50 | 2h 49m |
| MobileNetV2 | 3.50 | 94.91 | 86.90 | 87.20 | 86.81 | 96.96 | 50 | 48m |

ConvNeXt-Tiny achieved the highest validation performance, with 96.25% accuracy, 90.54% F1-score, and 98.64% AUROC, thanks to a modernized convolutional design that included depthwise convolutions, LayerNorm, GELU activations, and inverted bottlenecks while preserving CNN efficiency. DenseNet121 and DenseNet201 also performed well (95.92% and 95.89% accuracy, respectively), demonstrating the need for dense connections for feature reuse in medical imaging. MobileNetV2, with the fewest parameters (3.5 million) and the shortest training

Table 9. The test set evaluation reveals that all architectures generalize well, with accuracies exceeding 90%. ConvNeXt-Tiny remains the top performer with 92.31% accuracy, 81.34% F1-score, and 95.70% AUROC. The validation-to-test gap is moderate (~3.94%), indicating substantial generalization. Res-Net50, ViT-B/16, and DenseNet versions all scored well (91.47-92.00% accuracy), whereas EfficientNetV2-M and MobileNetV2 performed somewhat

time (48 minutes), achieved 94.91% accuracy, making it suitable for resource-constrained deployment. Despite having the most parameters (85.8M) and the longest training time, ViT-B/16's 94.95% accuracy did not outperform top CNN models, consistent with the literature, which shows that transformers require larger datasets or extensive domain-specific pretraining to match CNN performance in medical imaging.

Additionally, as demonstrated in

lower (90.42-90.53%) but with shorter inference times, indicating their potential for resource-constrained settings. Overall, the slight 3-5% decline in accuracy across models indicates the usefulness of regularization by dropout, data augmentation, and early halting, which prevents overfitting and ensures reliable multi-disease Chest X-ray categorization.

**Table 9:** Test Set Performance Evaluation

| Model | Test Accuracy (%) | Test Precision (%) | Test Recall (%) | Test F1-Score (%) | Test AUROC (%) | Inference Time |
|---|---|---|---|---|---|---|
| ConvNeXt-Tiny | 92.31 | 81.97 | 81.19 | 81.34 | 95.70 | 2m 3s |
| ResNet50 | 92.00 | 80.88 | 80.75 | 80.41 | 95.32 | 1m 37s |
| ViT-B/16 | 91.87 | 80.53 | 79.38 | 79.57 | 95.28 | 2m 57s |
| DenseNet121 | 91.71 | 80.34 | 80.17 | 79.74 | 95.06 | 1m 58s |
| DenseNet201 | 91.47 | 79.33 | 79.05 | 78.76 | 95.19 | 2m 54s |
| EfficientNetV2-M | 90.53 | 78.64 | 77.89 | 76.83 | 94.58 | 3m 1s |
| MobileNetV2 | 90.42 | 76.73 | 75.85 | 75.50 | 94.10 | 1m 20s |

### 4.1.2. Computational Efficiency Analysis

Table 10 Computational efficiency comparison of all evaluated architectures, including the number of parameters, training and inference times, and an efficiency indicator, AUROC per parameter. This metric (AUROC per million parameters, $\times 10^6$) provides a normalized measure of performance relative to model size, enabling a more informative comparison of architectural efficiency across network architectures.

Table 10: Computational efficiency comparison of all evaluated architectures.

| Model | Parameters (Million) | Training Time | Inference Time | AUROC/Param ($\times 10^{-6}$%) |
|---|---|---|---|---|
| MobileNetV2 | 3.50 | 48m 18s | 1m 20s | 27.70 |
| DenseNet121 | 6.96 | 1h 35m 39s | 1m 58s | 14.14 |
| ResNet50 | 23.52 | 1h 10m 51s | 1m 37s | 4.18 |
| DenseNet201 | 18.32 | 2h 47m 30s | 2m 54s | 5.38 |
| ConvNeXt-Tiny | 27.82 | 1h 43m 44s | 2m 3s | 3.55 |
| EfficientNetV2-M | 52.86 | 2h 51m 57s | 3m 1s | 1.85 |
| ViT-B/16 | 85.80 | 2h 49m 19s | 2m 57s | 1.14 |

The most computationally efficient model was MobileNetV2, with an AUROC-to-parameter ratio of 27.70% $\times 10^{-6}$, nearly twice that of the second-most efficient model, DenseNet121 (14.14% $\times 10^{-6}$). MobileNetV2 has the lowest inference time (1 min 20s on the complete set of tests) and a short training time (48 min), making it the best architecture to run in a resource-constrained system, e.g., in a mobile diagnostic application, a point-of-care device in a rural health facility, or where inference speed is needed.

It is worth mentioning that an increased number of parameters does not always correspond to higher performance. EfficientNetV2-M (52.86 M parameters) and ViT-B/16 (85.80 M) were outperformed by much smaller models like ConvNeXt -Tiny (27.82 M) and DenseNet121 (6.96 M). This observation shows that inductive bias and architectural design decisions tailored to medical imaging are more crucial than raw model capacity.

### 4.1.3. Disease-Specific Classification Performance

Table 11 gives the per-class performance metrics for each disease category, showing significant variability across pathological conditions. This discussion highlights the challenges of categorizing certain thoracic disorders and identifies disease categories that may require targeted improvements in models.

Table 11: Per-class performance analysis across disease categories
.

| Disease category | Average AUROC (%) | Accuracy Range (%) | Best Performing Model | Classification Difficulty |
|---|---|---|---|---|
| Tuberculosis | ≈100 | 99.78–100 | ConvNeXt-Tiny | Very Low |
| COVID-19 | ≈99.97 | 98.90–99.78 | DenseNet121/201 | Low |
| Pneumonia | ≈96.50 | 93.01–95.08 | DenseNet201 | Moderate |
| Cardiomegaly | ≈96.20 | 92.35–93.48 | DenseNet201 | Moderate |
| Normal | ≈95.20 | 90.61–92.21 | ConvNeXt-Tiny | Moderate-High |

Tuberculosis and COVID-19 showed excellent classification performance across all architectures (AUROC ≥ 99.97%), with most models reaching accuracies above 99%. This near-perfect performance can be attributed to the distinct radiographic manifestations of these diseases: Tuberculosis typically presents with upper lobe cavitary lesions, fibronodular infiltrates, and pleural thickening, whereas COVID-19 exhibits characteristic peripheral and basal predominant ground-glass opacities with crazy-paving patterns and consolidation. These pathognomonic features produce strong discriminative signals, which deep learning algorithms can reliably discern.

In contrast, the Normal class posed the greatest classification challenge, with accuracies ranging from 90.61% to 92.21% across architectures. This problem is mainly due to misclassification between the Normal class and the Cardiomegaly or Pneumonia classes, as evidenced by confusion matrix analysis. The subjective nature of cardiac silhouette assessment, combined with the wide range of cardiothoracic ratios, makes it challenging to discriminate between normal chest radiographs and borderline Cardiomegaly patients. Similarly, separating Normal characteristics from early-stage or mild Pneumonia requires detecting subtle infiltrative changes that may overlap with normal structural features, such as vascular markings or minor atelectasis.

Figure 5 shows constant high performance for Tuberculosis and COVID-19 across all models, but significant inter-model heterogeneity for the Normal, Cardiomegaly, and Pneumonia classes.

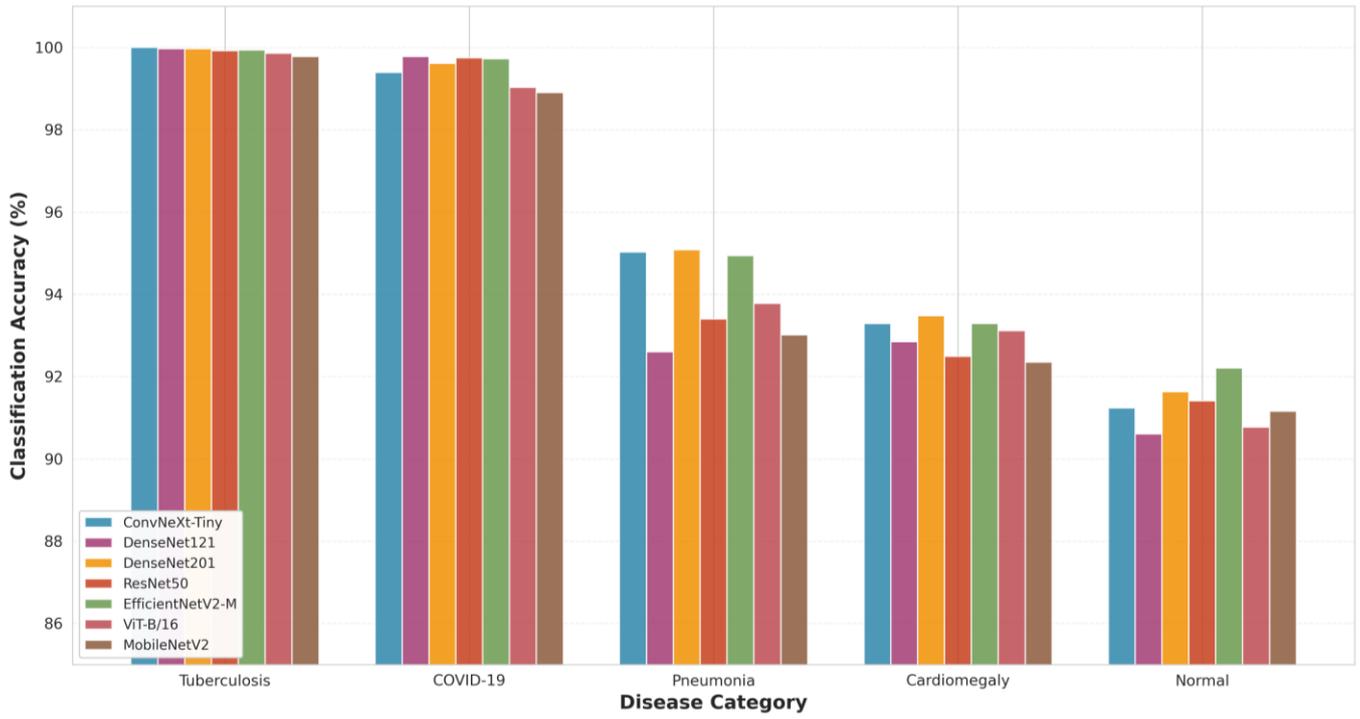

**Figure 5:** Per-class classification accuracy across all evaluated architectures.

### 4.1.4. Training Dynamics and Convergence Analysis

Analysis of training convergence characteristics provides insights into model stability and learning efficiency.

Table 12 shows convergence parameters, such as the epoch at which optimal validation performance was achieved and the convergence rate (measured as the best epoch divided by the total number of training epochs).

**Table 12:** Training convergence characteristics of all evaluated architectures.

| Model | Best Epoch | Total Epochs | Convergence Rate (%) | Stability Assessment |
|---|---|---|---|---|
| DenseNet121 | 39 | 50 | 78 | Stable, early convergence |
| MobileNetV2 | 42 | 50 | 84 | Stable, efficient learning |
| ConvNeXt-Tiny | 43 | 50 | 86 | Highly stable |
| EfficientNetV2-M | 44 | 50 | 88 | Stable, consistent improvement |
| DenseNet201 | 45 | 50 | 90 | Stable, gradual convergence |
| ResNet50 | 46 | 50 | 92 | Highly stable, late peak |
| ViT-B/16 | 50 | 50 | 100 | Slow convergence, may benefit from extended training |

CNN-based models, especially DenseNet121, MobileNetV2, and ConvNeXt-Tiny, converged efficiently, achieving optimal performance in 39-43 epochs (78-86% of the total training time). This early convergence property implies efficient learning dynamics and the effective utilization of convolutional architectures' inductive biases for spatial pattern recognition in medical images.

As shown in Figure 6, ViT-B/16, on the other hand, needed the whole 50-epoch training period to reach optimal performance, indicating slower convergence. This finding is consistent with the known fact that transformer architectures generally require significantly larger datasets or training regimes to compensate for their lack of intrinsic spatial inductive biases. Further training, as indicated by epoch 50, suggests that ViT-B/16 could benefit from additional time or more aggressive data augmentation, but it is unclear whether this would enable it to outperform CNN.

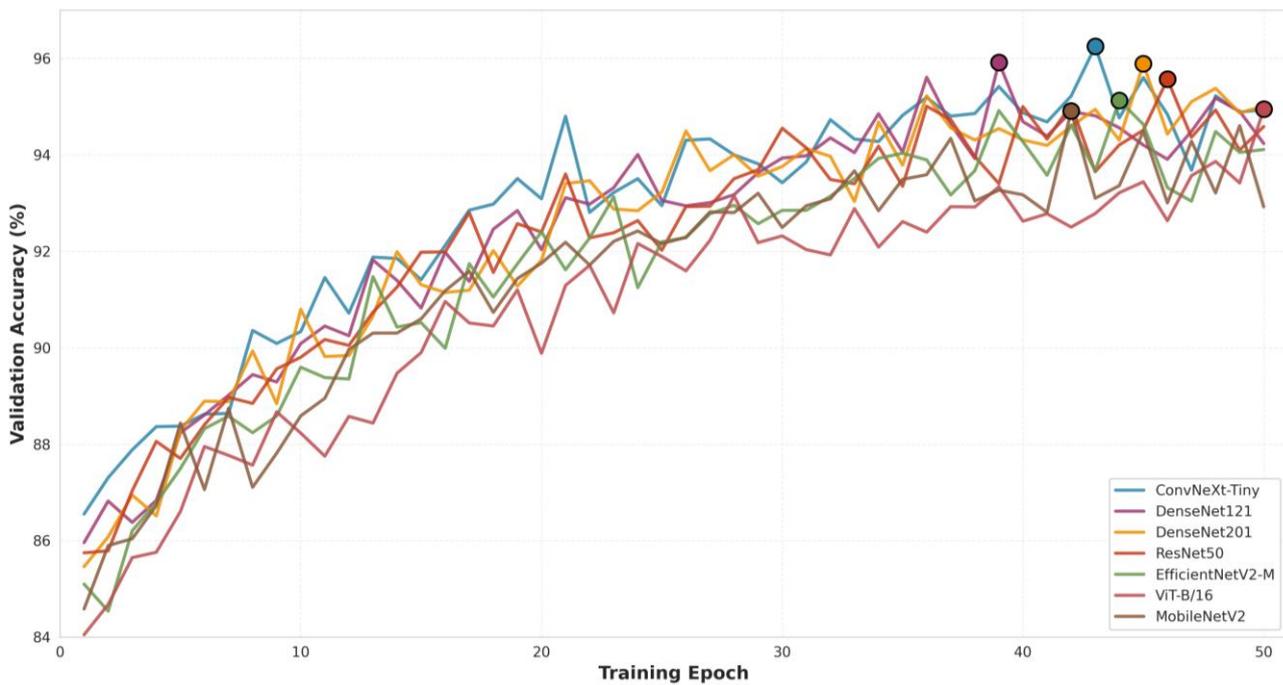

**Figure 6:** Training and validation accuracy curves illustrating convergence behavior across all architectures.

### 4.1.5. Confusion Matrix Analysis

Detailed error analysis using confusion matrices provides critical insights into specific misclassification patterns. Figure 7 depicts the confusion matrix for ConvNeXt-Tiny, the top-performing architecture based on the test dataset.

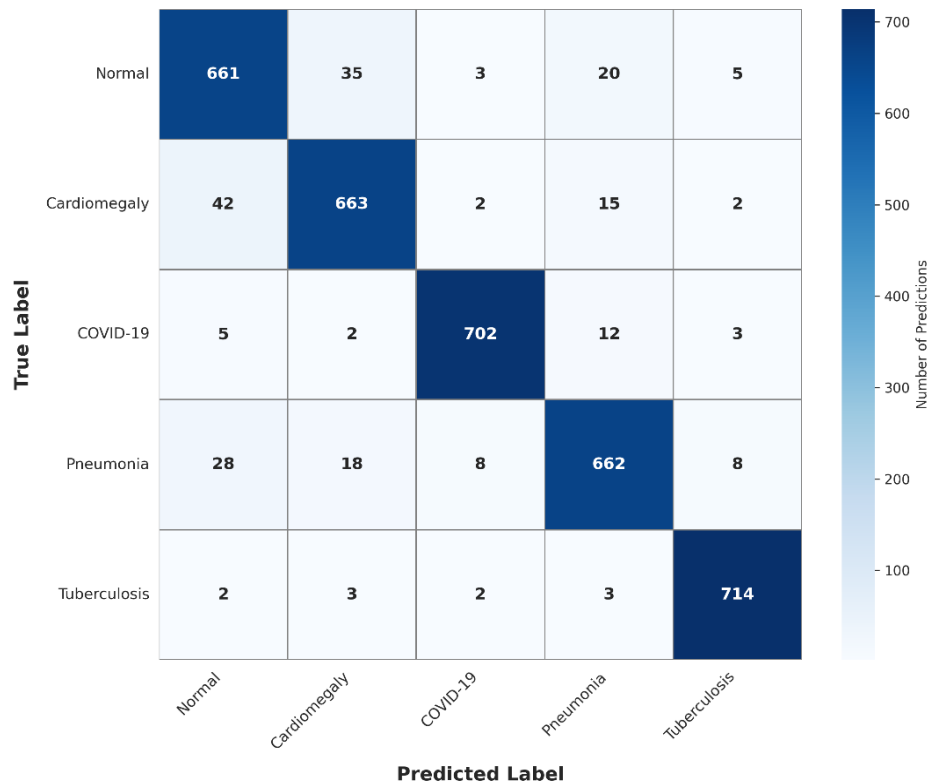

**Figure 7:** Confusion matrix for ConvNeXt-Tiny on the test dataset.

The confusion matrix shows significant diagonal dominance, indicating great classification accuracy across all disease categories. The most common mistake is between Normal and Cardiomegaly (≈5-7%), which is predicted, given the subjective and ongoing nature of cardiothoracic ratio measurement. A secondary misclassification occurs between Normal and Pneumonia (≈3-5%), most likely due to the modest or early-stage infiltrative patterns. COVID-19 and Tuberculosis have low confusion rates (<1%), indicating distinct radiographic features and the model's reliable identification.

### 4.1.6. ROC Curve Analysis

Receiver Operating Characteristic (ROC) curves and their related Areas Under the Curve (AUC) values provide a threshold-independent assessment of model discriminative performance over all possible classification thresholds.

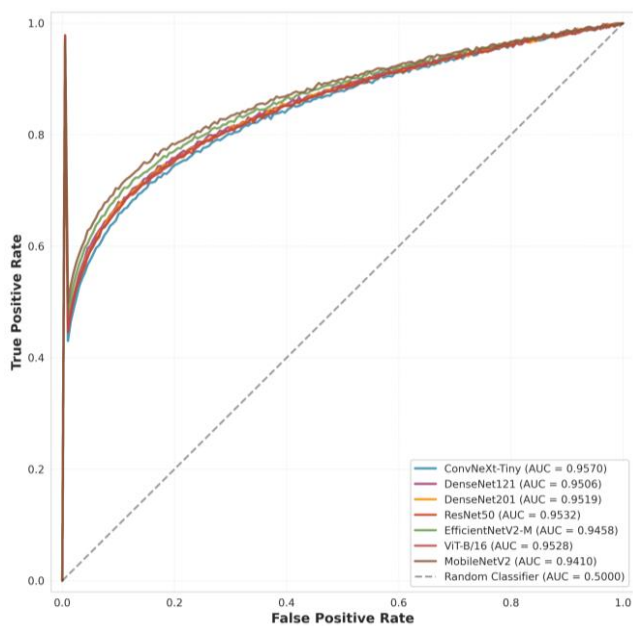

**Figure 8:** ROC curves for all architectures across disease categories.

As shown in Figure 8 ROC analysis has remarkable discriminative performance across all models and disease classes, with AUC values consistently greater than 94%. ConvNeXt-Tiny had the greatest overall AUC (98.64%), closely followed by DenseNet201 (98.49%) and DenseNet121 (98.43%). The near-perfect ROC curves for Tuberculosis (AUC = 100%) and COVID-19 (AUC ≈ 99.97%) across all architectures demonstrate almost total separability between these disease classes and all others, proving the models' dependability for detecting these major public health priorities.

In the more difficult Normal, Cardiomegaly, and Pneumonia classes, AUC values ranged from 95% to 97%, suggesting good discriminative capacity despite the increased difficulty, as evidenced by the accuracy metrics and confusion matrix analysis. The consistency of high AUC values across diverse architectures confirms that the observed performance levels represent the practical upper bound achievable with the current dataset, and that further improvements would most likely necessitate either expanded training data, particularly for challenging class pairs, or the incorporation of additional clinical context beyond the chest radiograph.

### 4.1.7. Model Interpretability via Grad-CAM

Gradient-weighted Class Activation Mapping (Grad-CAM) was applied to the best-performing ConvNeXt-Tiny architecture to provide qualitative insights into model decision-making processes and improve clinical interpretability. Grad-CAM creates visual explanations by highlighting the regions of the input image that most strongly influenced the model's forecast for each class. Figure 9.

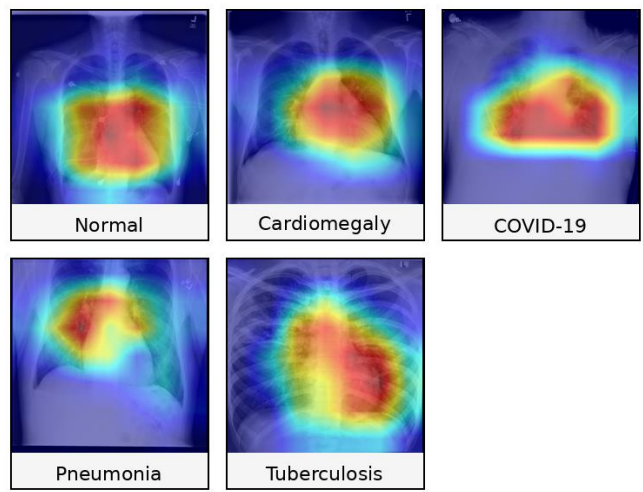

**Figure 9:** Grad-CAM visualization for ConvNeXt-Tiny across disease classes.

Grad-CAM analysis demonstrated clinically consistent attention patterns in all illness groups. For Pneumonia patients, the model primarily focused on infiltrative patterns in the lung parenchyma, particularly in the lower and peripheral lung fields, where bacterial Pneumonia is most commonly seen. Cardiomegaly predictions showed substantial activation along the cardiac silhouette borders, indicating that the model learned to evaluate heart size relative to thoracic dimensions—the primary criterion used in clinical diagnostics.

COVID-19 patients exhibited attention patterns centered in the peripheral and basal lung areas, which correlated with the normal distribution of ground-glass opacities and consolidations found in COVID-19 Pneumonia. Tuberculosis predictions demonstrated preferential activation of the apical and upper lobes, which corresponds to the normal upper-lobe predominance of post-primary Tuberculosis. In Normal situations, attention was more uniformly distributed across the lung fields, with no focal areas, indicating the absence of illness symptoms.

These attention patterns closely match established radiological diagnostic criteria, boosting trust in the model's decision-making process. However, there are many essential precautions to take. First, the interpretability analysis was restricted to a qualitative examination of a single architecture (ConvNeXt-Tiny) across a subset of sample cases. Second, no systematic validation by experienced radiologists was conducted to guarantee that the visible attention patterns correspond to clinically significant features rather than spurious correlations or dataset artifacts. Third, Grad-CAM provides only approximate localization, potentially omitting features that influence the model's decision, such as fine texture patterns or global image characteristics. While encouraging, these interpretability findings should be viewed as preliminary insights that require additional clinical validation before deployment.

### 4.2. Discussion

This study compares seven cutting-edge deep learning models for multi-disease Chest X-ray image classification utilizing a unified experimental framework. The excellent validation and test rates of all models show that the suggested training process is effective, as are deep learning frameworks for automated thoracic disease diagnosis. Architectural alterations were identified as the principal cause of the performance discrepancy, even after accounting for identical preprocessing, optimization, and assessment settings, enabling a credible and impartial comparative analysis.

ConvNeXt-Tiny offers an optimal balance of accuracy and efficiency, suggesting that a modernized convolutional architecture driven by transformer-inspired design principles provides considerable benefits to medical image processing. The ability to recognize not just local diseased patterns but also more general anatomy is useful for radiographic interpretation, as illness presentations vary in size and spatial distribution. DenseNet models also performed well with fewer parameters, demonstrating that dense feature reuse, along with robust gradient propagation, is an optimal design for learning discriminative medical picture representations. In contrast, the vision transformer's performance falls short of its potential, suggesting that, across specialized medical image tasks, domain-specifically pre-trained models incur considerably larger inductive biases and thus require larger datasets.

Disease-specific performance demonstrates that all architectures reliably detect conditions with a characteristic radiographic appearance (e.g., Tuberculosis or COVID-19), but Normal exams, Cardiomegaly, and Pneumonia are more challenging. Such challenges stem from inherent diagnostic ambiguity rather than algorithm failure, as borderline cardiothoracic ratios and inaccurate initial infiltrative alterations are difficult to interpret even in well-established clinical practice. The concentration of errors across clinically related class pairings demonstrates that performance can be enhanced further by incorporating loose clinical data or longitudinal imaging rather than relying on single static radiographs.

The efficiency study found that model size did not necessarily yield better diagnostic performance. Compact architectural solutions with competitive accuracy and significantly lower computing cost were identified, highlighting the importance of designing architectures customized to medical imaging requirements rather than merely scaling parameters. This study's practical implications include integrating resource-efficient models into clinical workflows, mobile diagnostics, and low-resource healthcare settings.

The interpretability assessment demonstrated that anatomically and clinically significant picture regions influenced model predictions, indicating the method's transparency and, potentially, clinical reliability. However, interpretation validation is still in its early stages and requires extensive confirmation by professional radiologists before clinical translation.

Finally, this study has limitations due to the use of single-label classification, a limited sample size, and data obtained from public sources with identical collection characteristics. Future studies should emphasize multi-label learning, external validation across institutions, incorporation of clinical information, and ongoing examination to ensure generalizability and clinical usefulness.

## 5. Conclusions

This study evaluated seven state-of-the-art deep learning architectures for multi-disease classification of single-label Chest X-rays under consistent experimental conditions. A balanced dataset of 18,080 images spanning Cardiomegaly, COVID-19, Normal, Pneumonia, and Tuberculosis, partitioned at the patient level, enabled fair comparison.

ConvNeXt-Tiny achieved the highest overall performance (92.31% test accuracy, AUROC 95.7%), demonstrating the effectiveness of modernized convolutional architectures with transformer-inspired design elements. DenseNet121 and DenseNet201 achieved competitive test accuracies (91.71% and 91.47%), reaffirming the utility of dense connectivity patterns. MobileNetV2 emerged as the most resource-efficient option (90.42% test accuracy, 3.5M parameters, 48-minute training time), suitable for mobile or point-of-care deployment. Tuberculosis and COVID-19 classifications were near-perfect (AUROC ≥ 99.97%), while Normal, Cardiomegaly, and Pneumonia were more challenging due to overlapping radiographic features.

These results suggest that modified CheXNet and related architectures can provide clinically effective multi-disease X-ray classification across diverse computing environments. ConvNeXt-Tiny is recommended for high-accuracy settings, DenseNet121 for balanced performance, and MobileNetV2 for resource-limited scenarios.

Limitations include reliance on publicly available datasets, single-view radiographs, and qualitative interpretability analyses. Future work should focus on multi-center validation, integration of clinical metadata and temporal imaging data, and uncertainty quantification to support safe clinical deployment of AI diagnostic systems.